\pdfoutput=1
\RequirePackage{ifpdf}
\ifpdf 
\documentclass[pdftex]{sigma}
\else
\documentclass{sigma}
\fi

\numberwithin{equation}{section}

\newtheorem{Theorem}{Theorem}[section]
\newtheorem*{Theorem*}{Theorem}
\newtheorem{Corollary}[Theorem]{Corollary}
\newtheorem{Lemma}[Theorem]{Lemma}
\newtheorem{Proposition}[Theorem]{Proposition}

\theoremstyle{definition}

\newtheorem{Example}[Theorem]{Example}
\newtheorem{Remark}[Theorem]{Remark}

\begin{document}
\allowdisplaybreaks

\newcommand{\arXivNumber}{2509.26243}

\renewcommand{\PaperNumber}{063}

\FirstPageHeading

\ShortArticleName{Symmetric Quantum Walks on Hamming Graphs and Their Limit Distributions}

\ArticleName{Symmetric Quantum Walks on Hamming Graphs\\ and Their Limit Distributions}

\Author{Robert GRIFFITHS~$^{\rm a}$ and Shuhei MANO~$^{\rm b}$}

\AuthorNameForHeading{R.~Griffiths and S.~Mano}

\Address{$^{\rm a)}$~School of Mathematics, Monash University, Clayton, Victoria 3800, Australia}
\EmailD{\mail{bob.griffiths@monash.edu}}

\Address{$^{\rm b)}$~The Institute of Statistical Mathematics, Tachikawa, Tokyo 190-8562, Japan}
\EmailD{\mail{smano@ism.ac.jp}}

\ArticleDates{Received November 07, 2025, in final form June 20, 2026; Published online July 03, 2026}

\Abstract{We study a class of symmetric coined quantum walks on Hamming graphs, where the distance between vertices specifies the transition probability. A special model is the simple quantum walk on the hypercube, which has been discussed in the literature. Eigenvalues of the unitary operator of the quantum walks are zeros of certain self-reciprocal polynomials. We obtain a spectral representation of the wave vector, where our systematic treatment relies on the coin space isomorphic to the state space and the commutative association scheme. The Grover coin is extended to the reflection about a vector in an invariant subspace of the Terwilliger algebra. The limit distributions of several quantum walks are obtained.}


\Classification{05E30; 33C45; 60B15; 60K40}

\section{Introduction}

Random walks on graphs are a typical research topic for
finite-state Markov chains. In particular, random walks
on cycles and on the hypercube are classical topics,
which can be viewed as random walks with the elements of
a finite group as their state space. See \cite{Dia88}
and references therein.

A simple random walk on the hypercube is a finite Markov
chain defined on a state space where each vertex on
the hypercube is binary-valued, with transitions occurring
only between adjacent vertices. If the dimension of
the hypercube is $d$, the state space can be viewed as
a~$d$-digit binary number, with transitions occurring only
between numbers that differ by only one digit. In this study,
we first discuss more general random walks on finite sets.
There are two directions of extension related to this
study: one extends the transition probability,
allowing transitions to occur between vertices that are
not necessarily adjacent. Such random walks are sometimes
called long-range random walks. There are several studies
on such random walks; see~\cite{CG21} and references there
in. The other extends the state space, defined on a state
space where each vertex on the hypercube is $n$-valued.
Here, two states that differ at only one vertex are called
adjacent. A graph in which adjacency is represented by
edges is called a Hamming graph.
The Hamming graph is a model of a word of length $d$
consisting of $n$ characters. The number of different
characters in two words is the distance in this paper,
and is called the Hamming distance. One motivation for
considering Hamming graphs is that they are
a fundamental object in string processing, including
coding theory \cite[Chapter~1]{BBIT21}.

Another motivation for considering Hamming graphs is
an interest in group representations and orthogonal
polynomials. The transition probability of a simple
random walk on the hypercube is invariant under
the action of the hyperoctahedral group, and therefore
its spectral representation is given by Krawtchouk
polynomials, associated with zonal spherical functions
\cite[Chapter~3]{Dia88}. Hora~\cite{Hor97}
established that the spectral representation of
the transition probability of a class of random walks
on the Hamming graph has Krawtchouk polynomial
eigenfunctions.

Quantum walks are motivated by search algorithms in
quantum computing. While not Markov chains, they are
models that incorporate randomness in the quantum
mechanical sense. A class of such models
is a coined, or discrete-time quantum walk on
finite graphs introduced by Aharonov et al.~\cite{AAKV01}.
They discussed quantum walks on cycles, and quantum
walks on various finite graphs have been explored in
the literature. However, coined quantum
walks are harder to analyse than random walks, and
explicit results are limited to a few models.

Within this context, quantum walks on the hypercube
have been the subject of study by numerous authors
since \cite{MR02}, and even the early paper \cite{SKW03},
which proposed using quantum walks for search,
considered quantum walks on the hypercube. However,
to the authors' knowledge, results concerning coined
quantum walks on the hypercube are limited to those
associated with the simple random walk, and there are
no results for general Hamming graphs. While results
for continuous-time quantum walks on Hamming graphs
exist \cite{BKMT08}, continuous-time quantum walks
are a significantly different model from coined
quantum walks, and their analysis is usually not as
hard as that of coined quantum walks.

Regarding the methods, for coined quantum walks on
the hypercube associated with the simple random walk,
\cite{MR02} and \cite{MPAD08} perform direct matrix
calculations, while \cite{SKW03} and subsequent papers
obtain results through quantum walks on the interval
defined by the distance. The latter has the advantage
of being able to use the results of the birth-death
process, as in \cite{HIKST18}. However, it seems
difficult to extend these methods to more general
quantum walks discussed in this paper.

In this paper, we discuss a class of symmetric coined
quantum walks
on Hamming graphs, where the distance between vertices specifies
the transition probability by using the commutative association
scheme. Section~\ref{sect:classic} introduces random walks on
the hypercube and Hamming graphs, on which the class of quantum
walks considered in this paper is based. While these results
are known, describing them using the commutative association
scheme prepares the stage for the next section.
Section~\ref{sect:quantum} describes the class of quantum walks
considered in this paper. We prepare a coin space isomorphic
to the state space and introduce the evolution operator.
The coin operator is the reflection about a vector in
an invariant subspace of the Terwilliger algebra.
We then give the Fourier transform. Section~\ref{sect:pol}
discusses the zeros of self-reciprocal polynomials. We show
that these zeros are aligned on the unit circle in the complex
plane, and provide an explicit form for special cases.
These zeros are eigenvalues of the evolution operator.
Section~\ref{sect:spectral} presents the main result of this
paper: the spectral representation of the wave vector
using the Krawtchouk polynomials. When each vertex is binary,
i.e., the hypercube, a particularly explicit form is obtained.
In Section~\ref{sect:limit}, as an application of
the spectral representation of the wave vector obtained
in Section~\ref{sect:spectral}, we give limiting distributions
for various quantum walks. We also reproduce known results for
quantum walks associated with simple random walks on the hypercube
as special cases.\looseness=-1

\section{The class of random walks}
\label{sect:classic}

We begin with a quick review of the Hamming graph and some
related notions. See \cite[Chapter~III]{BI84} and
\cite[Chapter~2]{BBIT21} for details.
Let a state space \smash{$X=\{0,1,\dots,n-1\}^d$} where $d,n\ge 2$.
Set $\partial(x,y)=|\{i\mid x_i\neq y_i\}|$ for
$x=(x_i), y=(y_i)$ $\in X$. The distance $\partial$ induces
a relation $X\times X$ by $(x,y)\in R_i$
$\Leftrightarrow$ $\partial(x,y)=i$, $i\in\{0,1,\dots,d\}$.
A pair of a finite set and a relation satisfying certain
conditions is called an association scheme, and
the association scheme \smash{$\bigl(X,\{R_i\}_{i\in\{0,\dots,d\}}\bigr)$}
introduced here is specifically called the Hamming scheme.
The advantage~of~for\-mu\-lat\-ing a~problem using an association
scheme is that it can reduce the need to perform direct
matrix calculations, as will be demonstrated throughout
this paper.

An undirected graph $(X,R_1)$ with vertices $X$ and edges $R_1$
is called a {\it Hamming graph} $H(d,n)$ (or the {\it hypercube}
if $n=2$). The adjacency matrix $A_i$ is defined by
\[
 (A_i)_{x,y}=
 \begin{cases}
 1 & \text{if } (x,y)\in R_i,\\
 0 & \text{if } (x,y)\notin R_i.
 \end{cases}
\]
Let $\mathcal{A}$ be the vector subspace of the matrices
$M_X(\mathbb{C})$ spanned by $A_0$, $A_1,\dots,A_d$.
Here, $\mathcal{A}$ is commutative with
the matrix product, and called a
Bose--Mesner algebra. Set
$\kappa_i=|\{{y\in X}\mid \partial(x,y)=i\}|$ (the right-hand
side being independent of $x\in X$), where
$\kappa_1$ is the degree of each vertex. Commutative $A_0$,
$A_1,\dots,A_d$ are simultaneously diagonalised by primitive
idempotents~$E_0$, $E_1,\dots,E_d$ in $\mathcal{A}$
satisfying $E_iE_j=E_i\delta_{i,j}$. Here
$E_0$ denotes the matrix whose entries are all~$1/n^d$.
The base change determines the coefficients $p_i(j)$ and
$q_i(j)$
\begin{equation}\label{eigenmat:1}
 A_i={\sum_{j=0}^d} p_i(j)E_j, \qquad
 n^dE_i=\sum_{j=0}^d q_i(j)A_j.
\end{equation}
In terms of the Krawtchouk polynomials
\begin{equation}\label{Kpol}
 K_i(j)=\sum_{l=\max\{0,i+j-d\}}^{\min\{i,j\}}(-1)^l(n-1)^{i-l}
 \begin{pmatrix}j\\l\end{pmatrix}
 \begin{pmatrix}d-j\\i-l\end{pmatrix},
\end{equation}
we have \cite[Section~III.2]{BI84}
\begin{equation}\label{eigenmat:2}
 p_i(j)=q_i(j)=K_i(j), \qquad \text{where} \quad
 \kappa_i=K_i(0)=(n-1)^i\begin{pmatrix}d\\i\end{pmatrix}.
\end{equation}
The generating function for the Krawtchouk polynomials is
\begin{equation*}
 \sum_{l=0}^dK_l(j)s^l=(1+(n-1)s)^{d-j}(1-s)^j.
\end{equation*}
Orthogonality is with respect to the Binomial distribution,
\[
 \sum_{l=0}^dK_i(l)K_j(l)
 \begin{pmatrix}d\\l\end{pmatrix}
 \left(1-\frac{1}{n}\right)^l\left(\frac{1}{n}\right)^{d-l}
 =\delta_{i,j}(n-1)^i
 \begin{pmatrix}d\\i\end{pmatrix}.
\]
Another version of Krawtchouk polynomials with different
normalization found in literature is~\cite[Section~1.10]{KS98}
\[
 \frac{K_i(j)}{\kappa_i}={}_2F_1(-i,-j;-d;n/(n-1)),
\]
which are the zonal spherical functions of the permutation
group $S_n \wr S_d$ on $X$.

Hora \cite{Hor97} discussed random walks on the Hamming graph
$H(d,n)$ with transition probability matrix $P$. He assumed
a spatial symmetry of $P$ that it is constant on each orbit
$R_i$:
\begin{equation}\label{assump:1}
 \partial(x,y)=\partial\bigl(x',y'\bigr) \quad \Rightarrow \quad
 (P)_{x,y}=(P)_{x',y'}
\end{equation}
or equivalently that $P$ belongs to Bose--Mesner algebra $\mathcal{A}$.

Under this assumption, the transition probability
takes the form of
\begin{equation}\label{trans}
 P=\sum_{i=0}^d\frac{w_i}{\kappa_i}A_i, \qquad
 \text{where} \quad w_i\ge 0,\ \sum_{i=0}^d w_i=1.
\end{equation}
Let $P_t(h)$, $t\in\mathbb{N}:=\{0,1,\dots\}$ denote
the $t$-step transition probability \smash{$\bigl(P^t\bigr)_{x,y}$} for
$(x,y)\in R_h$. Hora \cite{Hor97} established
a spectral representation
\begin{equation}\label{spec}
 P_t(h)=\frac{1}{n^d}\sum_{i=0}^d\rho_i^t K_i(h),
 \qquad h\in\{0,1,\dots,d\}
\end{equation}
with eigenvalues
\begin{equation}\label{eigen}
 \rho_i=\sum_{j=0}^d \frac{w_j}{\kappa_j} K_j(i)
 =\sum_{j=0}^d \frac{w_j}{\kappa_i} K_i(j),
 \qquad i\in\{0,1,\dots,d\}.
\end{equation}
This follows immediately. Since
\[
 P=\sum_{i=0}^d\frac{w_i}{\kappa_i}
 \sum_{j=0}^dp_i(j)E_j=\sum_{j=0}^d\rho_jE_j
\]
holds by~\eqref{eigenmat:1}, we have
\[
 P^t=\sum_{i=0}^d\rho_i^t E_i=
 \frac{1}{n^d}\sum_{i=0}^d\sum_{j=0}^d \rho_i^tq_i(j)A_j
 =\frac{1}{n^d}\sum_{i=0}^d\sum_{j=0}^d \rho_i^tK_i(j)A_j,
\]
where the first and last equalities follow
from $E_iE_j=E_i\delta_{i,j}$ and~\eqref{eigenmat:2},
respectively.
We note that $\rho_0=1$ and
\begin{equation}\label{bound}
 -\frac{1}{n-1}\le \rho_i\le 1, \qquad i\in\{1,\dots,d\}.
\end{equation}
For the lower bound in~\eqref{bound}, see \cite[Theorem~1]{DG12}. If a random walk is irreducible and aperiodic~$-1$
is not an eigenvalue. Since the transition probability
\eqref{trans} is symmetric, the stationary distribution
is uniform. Hora \cite{Hor97} gave a detailed treatment
of the cut-off phenomenon of a simple random walk
($w_i=\delta_{i,1}$). Collevechio and Griffiths \cite{CG21}
obtained~\eqref{spec} for a broad class of random walks on
the hypercube, i.e., $H(d,2)$, which contains the class
satisfying assumption~\eqref{assump:1}.

We consider a random walk starting from a state of $X$.
Without loss of generality, let the state be $0$.
In the standard basis of the vector space $\mathbb{C}^X$,
the element corresponding to $0$ is denoted by $e_0$.
Then,
\begin{equation}\label{trans_0}
 P_{0,x}=\sum_{i=0}^d \frac{w_i}{\kappa_i}(A_i)_{0,x}=
 \sum_{i=0}^d \frac{w_i}{\kappa_i}(A_ie_0)_x=
 \frac{w_{|x|}}{\kappa_{|x|}}, \qquad |x|:=\partial(0,x),
\end{equation}
where $\{(A_ie_0)_x:x\in X\}$ is the standard basis of
a $T$-invariant subspace, called the principal $T$-module, of
the Terwilliger algebra $T=\langle\mathcal{A},\mathcal{A}^*\rangle$
with respect to $e_0$, where $\mathcal{A}^*=\mathcal{A}^*(e_0)$
is the dual Bose--Mesner algebra of $\mathcal{A}$. Furthermore,
since $P\in \mathcal{A}$ by assumption~\eqref{assump:1}, we
can see that the state of the random walk is contained in
the principal $T$-module at any given time.
For the Terwilliger algebra, see \cite[Section~2.6]{BBIT21}.

\begin{Remark}
 The representation of the eigenvalues~\eqref{eigen}
 is known in classic Markov chain theory~\cite{DG12}.
 Suppose we have a Markov chain $\{Z_t \mid t\ge 0\}$
 whose stationary distribution is a~binomial
 distribution of length $d$ and parameter $1-1/n$,
 and the transition probability is
 \[
 P_{0,z}
 ={d\choose z}
 \left(1-\frac{1}{n}\right)^{z}\left(\frac{1}{n}\right)^{d-z}
 \sum_{i=0}^d\rho_iK_i(z).
 \]
 Then, $\rho_i$ has the representation~\eqref{eigen}.
 This Markov chain $\{Z_t\mid t\ge 0\}$ coincides with
 that of the distance $|x|$ with the transition
 probability~\eqref{trans} and starting from 0.
\end{Remark}

Some examples of random walks on $H(d,n)$ satisfying
assumption~\eqref{assump:1} are the following.

\begin{Example}[the simple random walk]\label{exam:simple}
 A walker at $x$ moves to a neighbour $y\in X$ satisfying
 $\partial(x,y)=1$ with equal probability,
 \[
 w_i=\delta_{i,1}, \qquad \rho_i=1-\frac{ni}{(n-1)d},
 \qquad i\in\{0,1,\dots,d\}.
 \]
 This random walk is irreducible and periodic if $n=2$
 and aperiodic otherwise.
\end{Example}

\begin{Example}[the independent random walk]\label{exam:indep}
 A walker at $x\in X$ moves to any $y\in X$ with equal
 probability,
 \[
 w_i=\frac{\kappa_i}{n^d}=
 \begin{pmatrix}d\\i\end{pmatrix}\frac{(n-1)^i}{n^d},
 \qquad \rho_i=\delta_{i,0}, \qquad i\in\{0,1,\dots,d\}.
 \]
 This random walk is irreducible and aperiodic. This
 random walk mixes in exactly one step.
\end{Example}

\begin{Example}[the non-local random walk with cardinality
 $m\in\{2,\dots,d\}$]\label{exam:nlrw}
 A walker at $x\in X$ moves to $y\in X$ satisfying
 $\partial(x,y)=m$ with equal probability,
 \[
 w_i=\delta_{i,m}, \qquad \rho_i=\frac{K_m(i)}{\kappa_m},
 \qquad i\in\{0,1,\dots,d\}.
 \]
 If $n=2$, this random walk is periodic, and irreducible
 if $m$ is odd and reducible otherwise. If~${n\ge 3}$,
 this random walk is irreducible and aperiodic.
\end{Example}

\begin{Example}[the mixture of i.i.d.\ updates for each coordinate]
 \label{exam:miid}
 The cardinality $i\!\in\!\{0,1,\dots,d\}$ is drawn
 from the binomial distribution of random parameter
 $\alpha\in(0,1)$ following some mixing measure.
 A walker at $x\in X$ moves to $y\in X$
 satisfying $\partial(x,y)=i$ with equal probability.
 Collevechio and Griffiths \cite{CG21} discussed
 the model of $n=2$.
 \[
 w_i=\begin{pmatrix}d\\i\end{pmatrix}\alpha^i(1-\alpha)^{d-i},
 \qquad \rho_i=
 \biggl(1-\frac{n\alpha}{n-1}\biggr)^i, \qquad
 i\in\{0,1,\dots,d\}.
 \]
 This random walk is irreducible and aperiodic.
\end{Example}

\section{The class of quantum walks}
\label{sect:quantum}

Let $X=\{0,1,\dots,n-1\}^d$ be position and coin space,
respectively, equipped with Hilbert spaces~$\mathcal{H}_P$
and~$\mathcal{H}_C$ with bases $\{|x\rangle\}_{x\in X}$
and $\{|y\rangle\}_{y\in X}$. The dual bases are denoted by
$\{\langle x|\}_{x\in X}$ and $\{\langle y|\}_{y\in X}$.
A quantum state at step $t\in\mathbb{N}$ is represented as
\[
 |\Psi(t)\rangle=\sum_{y\in X}\sum_{x\in X}\psi_{y,x}(t)|y,x\rangle,
 \qquad |y,x\rangle=|y\rangle\otimes|x\rangle,
\]
where $\{\psi_{y,x}(t)\}\in\mathbb{C}^{X\times X}$ is
called the wave vector. The probability that we observe
the quantum walker at position $x\in X$ after $t$-steps is
\[
 P_t(x)=\sum_{y\in X}|\psi_{y,x}(t)|^2.
\]
The evolution operator for one step of the quantum walk is
\begin{equation}\label{evo}
 U=S\circ(C\otimes I),
\end{equation}
where \smash{$C=\sum_{y,y'\in X}C_{y,y'}|y\rangle\langle y'|$} is called
a coin operator in $\mathcal{H}_C$, $I$ is the identity in
$\mathcal{H}_P$, and $S$ is the shift operator defined by
\[
 S=\sum_{y\in X}\sum_{x\in X}|y,x\oplus y\rangle\langle y,x|.
\]
Here, $x\oplus y$ is the component-wise sum of $d$-dimensional
vectors $x$ and $y$ of modulo $n$.
This shift operator reduces to the standard shift operator
in quantum walks on a hypercube associated with the simple
random walk used in \cite{MPAD08, MR02,SKW03}, when $n=2$ and $y$
is an element of the standard basis of the vector space
$\mathbb{C}^d$. Furthermore, $S^n=I$. Thus, $S$ is a natural
extension of the standard shift operator.
Applying $U$, we obtain the one-step transition of components
of the wave vector:
\begin{align}
 \psi_{y,x}(t+1)&=
 \sum_{y'\in X}\sum_{x'\in X}U_{y,x;y',x'}\psi_{y',x'}(t)
 \nonumber\\
 &=\sum_{y'\in X}C_{y,y'}
 {\psi_{y',x\oplus(n-1)y}}(t),
 \qquad x,y \in X,\ t\ge 0, \label{trans_q}
\end{align}
where
\[
 \oplus(n-1)y=
 \overbrace{\oplus y\oplus y\oplus y\oplus\cdots\oplus y}^{(n-1)\text{-times}}
\]
and
\[
 U=\sum_{y\in X}\sum_{y'\in X}\sum_{x\in X}\sum_{x'\in X}
 U_{y,x;y',x'}|y,x\rangle\langle y',x'|.
\]

We consider a class of coined quantum walks on Hamming
graph $H(d,n)$ stimulated by random walks discussed in
the previous section. As for the coin operator, we
take a common choice, so called Szegedy's walk
\cite{Szegedy04}, associated with the random
walk determined by the transition probability~\eqref{trans}.
Namely,
\eqref{trans_0} determines
\begin{equation}\label{coin}
 C_{y,y'}=2\sqrt{P_{0,y}P_{0,y'}}-\delta_{y,y'}
 =2\sqrt{\frac{w_{|y|}}{\kappa_{|y|}}\frac{w_{|y'|}}
 {\kappa_{|y'|}}}-\delta_{y,y'}, \qquad y,y'\in X.
\end{equation}
We can confirm that $C$ is orthogonal.
This coin operator reduces to the standard $d$-dimensional
Grover coin in quantum walks on a hypercube associated with
the simple random walk used in~\mbox{\cite{MR02,SKW03}}, when $n=2$
and $w_i=\delta_{i,1}$. Since
\[
 \sum_{y\in X}\sum_{y'\in X}
 \sqrt{\frac{w_{|y|}}{\kappa_{|y|}}\frac{w_{|y'|}}{\kappa_{|y'|}}}
 |y\rangle\langle y'|
\]
is a projection operator to the principal $T$-module, $C$ is
a reflection about a unit vector in the principal $T$-module with respect to $0$.
We set the initial state
\begin{equation}\label{init}
 \psi_{y,x}(0)=\delta_{x,0}\sqrt{\frac{w_{|y|}}{\kappa_{|y|}}},
 \qquad x,y\in X,
\end{equation}
which means that the quantum walk starts from position
$0$ with law~\eqref{trans_0} in the coin space
\[
 |\psi_{y,x}(0)|^2=\delta_{x,0}\frac{w_{|y|}}{\kappa_{|y|}},
 \qquad x,y\in X,
\]
where \smash{$\sum_{y\in X}\sum_{x\in X}|\psi_{y,x}(0)|^2=1$}, since
\smash{$\sum_{y\in X}w_{|y|}/\kappa_{|y|}=1$}.
Moreover, the initial coin
\[
 \sum_{y\in X}\sqrt{\frac{w_{|y|}}{\kappa_{|y|}}}|y\rangle
 =\sum_{y\in X}\sum_{i=0}^d\sqrt{\frac{w_i}{\kappa_i}}(A_ie_0)_y|y\rangle
\]
is the unit vector in the principal $T$-module.
An observation here is that
\smash{$\{\psi_{y,x}(t)\}\in\mathbb{R}^{X\times X}$} for all
$t\in \mathbb{N}$.

To solve~\eqref{trans_q}, we employ the Fourier transform
in the position space, which is
 standard in analyses of quantum walks on graphs \cite{AAKV01},
\[
 \tilde{\psi}_{y,\xi}(t)=\frac{1}{\sqrt{n^d}}
 \sum_{x\in X}\zeta^{\xi\cdot x}\psi_{y,x}(t)
\]
and the inverse transform
\[
 \psi_{y,x}(t)=\frac{1}{\sqrt{n^d}}
 \sum_{\xi\in X}\zeta^{-\xi\cdot x}\tilde{\psi}_{y,\xi}(t),
\]
where \smash{$\zeta\equiv {\rm e}^{2\pi\sqrt{-1}/n}$},
\smash{$\xi\cdot x=\sum_{i=1}^d\xi_i x_i$}. We have
\begin{align}
 \tilde{\psi}_{y,\xi}(t+1)&=
 \frac{1}{\sqrt{n^d}}\sum_{y'\in X}\sum_{x\in X}
 \zeta^{\xi\cdot x}C_{y,y'}\psi_{y',x\oplus(n-1)y}(t)
 =\frac{1}{\sqrt{n^d}}\sum_{y'\in X}\sum_{x'\in X}
 \zeta^{\xi\cdot(x'\oplus y)}C_{y,y'}\psi_{y',x'}(t)\nonumber\\
 &=\zeta^{\xi\cdot y}\sum_{y'\in X}C_{y,y'}\tilde{\psi}_{y',\xi}(t),
 \qquad y,\xi\in X,\ t\ge 0,
 \label{trans_qd}
\end{align}
where in the second equality we set {$x'=x\oplus(n-1)y$} and used
\smash{$\zeta^{\xi\cdot x}={\zeta^{\xi\cdot(x\oplus ny)}=\zeta^{\xi\cdot(x'\oplus y)}}$},
since~${\zeta^{n\xi\cdot y}=1}$.
The advantage of working in the $\xi$-coordinate is that
\eqref{trans_qd} becomes a system of equations that is separated
with respect to each $\xi$.
Although~\eqref{trans_qd} is not separated in the coin space,
it can be solved, as we will see in Section~\ref{sect:spectral}.

\section{Zeros of a self-reciprocal polynomial}
\label{sect:pol}

A polynomial $p_n(z)$ of degree $n$ is self-reciprocal if
$p_n(z)=p_n^*(z)$, \smash{$p_n^*(z):=z^n\overline{p_n(1/\bar{z})}$}.
The distribution
of the zeros of such a polynomial are interesting in
their own right. At the end of this section, we will see
that the zeros are the eigenvalues of the evolution
operator of the quantum walks. The following result is
anticipated because $U$ is unitary.

\begin{Lemma}\label{lemm:pol}
 For constant $\rho\in [-1/(n-1),1]$, $n\ge 2$, all of the zeros
 of polynomial
 \begin{equation}\label{pol}
 z^n+2\rho\sum_{i=1}^{n-1}z^i+1, \qquad z\in\mathbb{C}
 \end{equation}
 are on the unit circle, namely, $\{z\in\mathbb{C}\mid |z|=1\}$.
\end{Lemma}

\begin{proof}
 Suppose $1\ge \rho>0$. The polynomial~\eqref{pol} is
 self-reciprocal and represented as
 \[
 p_n(z)=z^n+2\rho\sum_{i=1}^{n-1}z^i+1=zq_{n-1}(z)+q_{n-1}^*(z),
 \]
 where \smash{$q_{n-1}(z)=z^{n-1}+\rho\sum^{n-2}_{i=0}z^i$}.
 According to Chen \cite[Theorem~1]{Che95}, all the zeros
 of $p_n(z)$ lie on the unit circle if all the zeros of
 $q_{n-1}(z)$
 are in or on the unit circle. By the Enestr\"om--Kakeya theorem,
 all the zeros of $q_{n-1}(z)$ lie in or on the unit circle
 since $1\ge \rho>0$. Therefore, the assertion holds.
 If $\rho=0$, the zeros are the $n$-th roots of $-1$.
 Finally, suppose $-1/(n-1)\le \rho<0$. Theorem~1 of
 Lakatos and Losonczi \cite{LL04} says that all the zeros of
 a self-reciprocal polynomial
 $\sum_{i=0}^na_iz^i$, $a_i=a_{n-i}$, $i\in\{0,\dots,n\}$
 are on the unit circle if
 \[
 |a_0|\ge\frac{1}{2}\sum_{i=1}^{n-1}|a_i|,
 \]
 and $p_n(z)$ satisfies this if $-1/(n-1)\le \rho\le 1/(n-1)$.
\end{proof}

In the following, we assume $n$ is
prime, namely, \smash{$\zeta^k={\rm e}^{2\pi\sqrt{-1}k/n}$},
$k\in\{1,2,\dots,n-1\}$ are the primitive roots of unity.
This assumption makes following expressions explicit.
We collect some properties of the zeros of the polynomial~\eqref{pol}.

\begin{Proposition}\label{prop:rho}
Assume $n$ is a prime. If unity is a zero of the polynomial~\eqref{pol},
 then ${\rho=-1/(n-1)}$, and if a primitive
 root of unity is a zero, then $\rho=1$.
\end{Proposition}

\begin{proof}
 The first assertion follows immediately. The second assertion
 follows by
 \[
 \sum_{i=1}^{n-1}\zeta^{ki}=\sum_{i=0}^{n-1}\zeta^{ki}-1
 =\frac{1-\zeta^{kn}}{1-\zeta^k}-1=-1,
 \qquad k\in\{1,\dots,n-1\},
 \]
 where $\zeta\equiv {\rm e}^{2\pi\sqrt{-1}/n}$.
\end{proof}

\begin{Proposition}
 For prime $n$ $(\ge 3)$, the zeros of the polynomial~\eqref{pol} are
 $-1$ and the following:
 \begin{itemize}\itemsep=0pt
 \item[$(i)$] if $\rho=1$, the primitive roots of unity
 $\zeta^k={\rm e}^{2\pi\sqrt{-1}k/n}$, $k\in\{1,2,\dots,n-1\}$;
 \item[$(ii)$] if $\rho=0$,
 $-\zeta^k$, $k\in\{1,2,\dots,n-1\}$;
 \item[$(iii)$] if $\rho=-1/(n-1)$, unity, and those on the unit
 circle except for $\pm 1$ and the primitive roots of unity
 if $n\ge 5$.
 \end{itemize}
\end{Proposition}

\begin{proof}
 ($i$) The polynomial factors as
 $(z+1)\bigl(z^{n-1}+z^{n-2}+\cdots+z+1\bigr)$. Since the second factor
 is the $n$-th cyclotomic polynomial, the zeros are the $n$-th
 primitive roots of unity. ($ii$) Similar to ($i$). ($iii$)
 The polynomial factors as
 \[
 (z-1)^2(z+1)\Biggl\{
 \sum_{i=1}^{n-2}
 \biggl\lceil\frac{i}{2}\biggr\rceil
 \biggl(n-2\biggl\lfloor\frac{i}{2}\biggr\rfloor-1\biggr)
 \frac{z^{n-i-2}}{n-1}\Biggr\}.
 \]
 The last factor is a polynomial of order $n-3$.
 The zeros of the polynomial are on the unit circle
 by Lemma~\ref{lemm:pol}, and not the primitive roots
 of unity by Proposition~\ref{prop:rho}
\end{proof}

In the following part of this paper, the zeros of the polynomial
\eqref{pol} with replacing $z$ by $-z$:%
\begin{equation}\label{pol-}
 (-z)^n+2\rho\sum_{i=1}^{n-1}(-z)^i+1, \qquad z\in\mathbb{C}
\end{equation}
appear.

\begin{Remark}
 If $n=2$,~\eqref{pol-} gives $2\rho=z+1/z$ and the real part
 of the zeros is $\rho$, since a zero is on the unit circle.
 The mapping $z\mapsto z+1/z$ is a conformal map known as
 the Joukowsky transform. The Joukowsky transform maps
 the unit circle to the real interval $[-2,2]$. For a~prime
 $n\ge 3$, we have
 \[
 2\rho=\frac{1}{\sum_{i=1}^{n-1}z^i}+\frac{1}{\sum_{i=1}^{n-1}z^{-i}}.
 \]
 The mapping \smash{$z\mapsto 1/\sum_{i=1}^{n-1}z^i+1/\sum_{i=1}^{n-1}z^{-i}$}
 is also a conformal map which maps the unit circle to the
 real interval. The argument $\theta$ of a zero satisfies
 \[
 2\rho
 =1+\frac{\cos\theta-\cos(n\theta)}{1-\cos\{(n-1)\theta\}}.
 \]
\end{Remark}

\begin{Proposition}\label{prop:eigen_q}
 Consider a random walk on a Hamming graph $H(d,n)$, $d\ge 2$
 and prime $n$ satisfying assumption~\eqref{assump:1}.
 Any eigenvalue of the evolution operator~\eqref{evo}
 of the quantum walk with coin operator~\eqref{coin}
 is a zero of the polynomial~\eqref{pol-} with
 $\rho=\rho_{|\xi|}$ for some $\xi\in X$.
\end{Proposition}

We prepare a lemma to prove Proposition~\ref{prop:eigen_q}.
It provides the Fourier transform of functions of
$|z|=\partial(0,z)$ by the characters of the direct
product of the cyclic groups of order $n$.

\begin{Lemma}\label{lemm:sum}
 For any function $f\colon \{0,1,\dots,d\}\to\mathbb{C}$
 and prime $n$, we have
 \begin{equation}\label{sum}
 \sum_{z\in X}\zeta^{k\xi\cdot z}f(|z|)
 =\sum_{j=0}^dK_j(|\xi|)f(j), \qquad \xi\in X,
 \quad k\in\{1,\dots,n-1\}
 \end{equation}
 and
 \begin{equation}\label{sum2}
 \sum_{\xi\in X}\zeta^{-k\xi\cdot z}f(|\xi|)
 =\sum_{j=0}^dK_j(|z|)f(j), \qquad z\in X,
 \quad k\in\{1,\dots,n-1\},
 \end{equation}
 where $\zeta\equiv {\rm e}^{2\pi\sqrt{-1}/n}$.
\end{Lemma}

\begin{proof}
 Since $\zeta^{jn}=1$ for $j\in\mathbb{Z}$,
 \begin{equation}\label{lemm:sum:1}
 \sum_{l=1}^{n-1}\zeta^{kjl}
 =\frac{1-\zeta^{kjn}}{1-\zeta^{kj}}-1=-1,
 \qquad j,k\in\{1,\dots,n-1\},
 \end{equation}
 where \smash{$\zeta^{kj}\!=\!{\rm e}^{2\pi \sqrt{-1}kj/n}\!\neq\! 1$} since $n$ is prime.
 Without loss of generality, we assume
${\xi_1,\dots,\xi_{|\xi|}\!>\!0}$
and $\xi_{|\xi|+1}=\cdots=\xi_{d}=0$.
 Fix $|z|\in\{0,\dots,d\}$, $l\in\{0,\dots,\min\{|\xi|,|z|\}\}$,
 and suppose $z_{i_1},\dots,z_{i_l}>0$,
 $\{i_1,\dots,i_l\}\in\{1,\dots,|\xi|\}$ and
 \smash{$z_{i_{l+1}},\dots,z_{i_{|z|}}>0$},
 \smash{$\bigl\{i_{l+1},\dots,i_{|z|}\bigr\}\in\{|\xi|\!+\!1,\dots,d\}$}.
 The contribution of such $z_1,\dots,z_d$ to the
 left-hand side of~\eqref{sum} is $f(|z|)$ times
 \begin{align}
 &\begin{pmatrix}|\xi|\\l\end{pmatrix}
 \begin{pmatrix}d-|\xi|\\|z|-l\end{pmatrix}
 \prod_{j=1}^l\Biggl(
 \sum_{z_{i_j}=1}^{n-1}\zeta^{k\xi_jz_{i_j}}
 \Biggr)
 (n-1)^{|z|-l}
 =\begin{pmatrix}|\xi|\\l\end{pmatrix}
 \begin{pmatrix}d-|\xi|\\|z|-l\end{pmatrix}
 (-1)^l(n-1)^{|z|-l}\label{lemm:sum:2},
 \end{align}
 where we used~\eqref{lemm:sum:1}.
 Summing up~\eqref{lemm:sum:2} in $l$ yields
 \begin{equation}\label{lemm:sum:3}
 \sum_{l=\max\{0,|z|+|\xi|-d\}}^{\min{\{|\xi|,|z|}\}}
 \begin{pmatrix}|\xi|\\l\end{pmatrix}
 \begin{pmatrix}d-|\xi|\\|z|-l\end{pmatrix}
 (-1)^l(n-1)^{|z|-l}=K_{|z|}(|\xi|).
 \end{equation}
 Summation of~\eqref{lemm:sum:3} in $|z|$ is
 the right-hand side of~\eqref{sum}.
 We can confirm~\eqref{sum2} in the same manner.
\end{proof}

\begin{Remark}
 Consider the map $\eta_j\colon x_j\mapsto \zeta^{\xi_jx_j}$, where
 \smash{$\eta(x)=\prod_{j=1}^d\eta_j(x_j)=\zeta^{\xi\cdot x}$}
 comprises the character group of the direct product of
 the cyclic group of order $n$. If we set
 $f(z)=\delta_{z,j}$ and~${k=1}$, then~\eqref{sum} reduces
 to the relation
 \begin{equation}\label{schur}
 \eta(X_j)=\sum_{x\in X\colon |x|=j}\eta(x)
 =\sum_{x\in X\colon |x|=j}\zeta^{\xi\cdot x}=K_j(|\xi|),
 \end{equation}
 where $X_j=\sum_{x\in X\colon |x|=j}x$ is an element of
 a Schur-ring over
 $X$. This relation is Proposition~2.2 in Section~III,
 \cite{BI84} and used to establish~\eqref{eigenmat:2}.
 However, note that~\eqref{schur} holds for any $n\ge 2$,
 while $n$ should be prime for~\eqref{sum} if
 $k\in\{2,\dots,n-1\}$.
\end{Remark}

An immediate consequence for~\eqref{eigen} is the following.

\begin{Corollary}
 We have
 \begin{equation}\label{eigen_ftr}
 \rho_{|\xi|}=\sum_{z\in X}\zeta^{k\xi\cdot z}
 \frac{w_{|z|}}{\kappa_{|z|}}, \qquad \xi\in X,
 ~k\in\{1,\dots,n-1\}.
 \end{equation}
\end{Corollary}

\begin{proof}[Proof of Proposition~\ref{prop:eigen_q}]
 Let
 \smash{$|v\rangle=\sum_{y\in X}\sum_{x\in X} v_{y,x}|y,x\rangle$}
 and $\mu$ be an eigenvector and the eigenvalue of
 the evolution operator~\eqref{evo}, respectively.
 That is, we have
 \begin{equation}\label{eigen_q}
 \sum_{y'\in x}\sum_{x'\in X}U_{y,x;y',x'}v_{y',x'}
 =\sum_{y'\in X}C_{y,y'}v_{y',x\oplus(n-1)y}
 =\mu v_{y,x}.
 \end{equation}
 Let
 \[
 u_{y,\xi}=\sqrt{\frac{w_{|y|}}{n^d\kappa_{|y|}}}
 \sum_{x\in X}\zeta^{\xi\cdot x}v_{y,x}
 \]
 for some $\xi\in X$. We consider the following two cases.

\emph{Case $1$.} Suppose we can take $\xi$ such that
 $\sum_{y\in X} u_{y,\xi}\neq 0$.
 By the same argument as obtaining~\eqref{trans_qd} from~\eqref{trans_q}, we recast the right
 equality of~\eqref{eigen_q} into
 \begin{equation}\label{prop:eigen_q:1}
 \mu\zeta^{-\xi\cdot y}u_{y,\xi}=-u_{y,\xi}
 +2\frac{w_{|y|}}{\kappa_{|y|}}\sum_{y'\in X}u_{y',\xi}.
 \end{equation}
 Summing up both sides of~\eqref{prop:eigen_q:1} in $y\in X$
 gives
 \begin{equation}\label{prop:eigen_q:2}
 \mu\sum_{y\in X}\zeta^{-y\cdot\xi}u_{y,\xi}=
 -\sum_{y\in X}u_{y,\xi}+2\sum_{y\in X}\frac{w_{|y|}}{\kappa_{|w|}}
 \sum_{y'\in X}u_{y',\xi}=\sum_{y\in X}u_{y,\xi},
 \end{equation}
 because \smash{$\sum_{y\in X}w_{|y|}/\kappa_{|y|}=1$}.
 Multiplying by \smash{$\zeta^{ky\cdot\xi}$},
 $k\in\{1,\dots,n-1\}$ and summing up both sides of~\eqref{prop:eigen_q:1} gives
 \begin{equation}\label{prop:eigen_q:3}
 \mu\sum_{y\in X}\zeta^{(k-1)y\cdot\xi}u_{y,\xi}=
 -\sum_{y\in X}\zeta^{ky\cdot\xi}u_{y,\xi}
 +2\rho_{|\xi|}\sum_{y\in X}u_{y,\xi},
 \end{equation}
 where we used~\eqref{eigen_ftr}.
 Since $\zeta^{nz\cdot\xi}=1$, recursive use of~\eqref{prop:eigen_q:3}
 gives
 \begin{align*}
 \mu^n\sum_{y\in X}u_{y,\xi}&=
 -\mu^{n-1}\sum_{y\in X}\zeta^{y\cdot\xi}u_{y,\xi}
 +2\mu^{n-1}\rho_{|\xi|}\sum_{y\in X}u_{y,\xi}\\
 &=\mu^{n-2}\sum_{y\in X}\zeta^{2y\cdot\xi}u_{y,\xi}
 -2\mu^{n-2}\rho_{|\xi|}\sum_{y\in X}u_{y,\xi}
 +2\mu^{n-1}\rho_{|\xi|}\sum_{y\in X}u_{y,\xi}\\
 &=(-1)^{n-1}\mu\sum_{y\in X}\zeta^{(n-1)y\cdot\xi}u_{y,\xi}
 +2\rho_{|\xi|}\sum_{j=1}^{n-1}(-1)^{n-j-1}\mu^{j}\sum_{y\in X}u_{y,\xi}\\
 &=(-1)^{n-1}\sum_{y\in X}u_{y,\xi}
 +2\rho_{|\xi|}\sum_{j=1}^{n-1}(-1)^{n-j-1}\mu^{j}\sum_{y\in X}u_{y,\xi},
 \end{align*}
 where~\eqref{prop:eigen_q:2} is used in the last equality.
 Since \smash{$\sum_{y\in X}u_{y,\xi}$} is non-zero, we have{\samepage
 \[
 (-\mu)^n+2\rho_{|\xi|}\sum_{j=1}^{n-1}(-\mu)^j+1=0,
 \]
 which shows that $\mu$ is a zero of the polynomial~\eqref{pol-}.}

\emph{Case $2$.} Suppose $\sum_{y\in X} u_{y,\xi}=0$ for all
 $\xi\in X$. Since
 \[
 \sum_{y\in X}\sqrt{\frac{w_{|y|}}{\kappa_{|y|}}}v_{y,x}=
 \sum_{y\in X}\sum_{\xi\in X}\zeta^{-\xi\cdot x}u_{y,\xi}
 \sum_{\xi\in X}\zeta^{-\xi\cdot x}\sum_{y\in X}u_{y,\xi}
 =0,
 \]
 the vector \smash{$\sum_{y\in X} v_{y,x}|y\rangle$} is in the orthogonal
 complement of the principal $T$-module, and~\eqref{eigen_q}
 reduces to $v_{y,x\oplus(n-1)y}=-\mu v_{y,x}$.
 Therefore, we have
 \[
 (-\mu)^n v_{y,x}=(-\mu)^{n-1}v_{y,x\oplus(n-1)y}=\cdots=v_{y,x\oplus n(n-1)y}=v_{y,x}
 \]
 for all $x,y\in X$. If $n=2$, $\mu=1$ or $-1$, which
 are the zeros of the polynomial~\eqref{pol-} with
 $\rho=-1$ and $\rho=1$, respectively. If $n\ge 3$,
 $-\mu$ is an $n$-th root of unity, and by Proposition~\ref{prop:rho},
 it is a~zero of the polynomial~\eqref{pol-} with
 $\rho=-1/(n-1)$ or $\rho=1$.
\end{proof}

\section{Spectral representation of wave vectors}
\label{sect:spectral}

The following spectral representations of the wave vector
of the quantum walks on Hamming graphs are the main results
of this paper. In this section, we establish them.

\begin{Theorem}[spectral representation of wave vector]
 \label{theo:n>2}
 Consider a random walk on a Hamming graph $H(d,n)$,
 $d\ge 2$ and prime $n$ satisfying
 assumption~\eqref{assump:1} with the eigenvalues
 ${-1/(n-1)<\rho_j<1}$, $j\in\{1,\dots,d\}$. Let
 \smash{$\mu^{(1)}_j,\dots,\mu^{(n)}_j$} be the zeros of
 the polynomial~\eqref{pol-} with $\rho=\rho_j$,
 $j\in\{1,\dots,d\}$ and assume they are distinct
 for each $j$. The wave vector of the quantum walk
 with coin operator~\eqref{coin} and the initial state~\eqref{init}
 is represented as
 \begin{align*}
 {\psi}_{y,x}(t)={}&
 \frac{1}{n^d}\sqrt{\frac{w_{|y|}}{\kappa_{|y|}}}
 \sum_{k=0}^{n-1}\sum_{l=0}^{n-1}
 \frac{\zeta^{lk}}{n}\nonumber\\
 &{\times} \Biggl\{1+\sum_{j=1}^d K_j(|x\oplus ly|)
 \Biggl[\bigl(-\zeta^{-k}\bigr)^t
 \Biggl(1-\sum_{i=1}^n\frac{2c_{j}^{(i)}}{1+\zeta^k
 \mu_{j}^{(i)}}\Biggr)
 +\sum_{i=1}^n
 \frac{2c_{j}^{(i)}\bigl(\mu_{j}^{(i)}\bigr)^t}
 {1+\zeta^k\mu^{(i)}_{j}}
 \Biggr]\Biggr\}, 
 \end{align*}
 where
 \begin{equation}\label{coeff}
 c_j^{(i)}=\frac{\bigl(\mu_j^{(i)}\bigr)^n+(1-\rho_j)
 \bigl(\mu_j^{(i)}\bigr)^{n-1}
 -\rho_j(-1)^n}
 {n\bigl(\mu_j^{(i)}\bigr)^n+[n-2\rho_j(n-1)]\bigl(\mu_j^{(i)}\bigr)^{n-1}
 -2\rho_j(-1)^n\sum_{k=0}^{n-2}\bigl(-\mu_j^{(i)}\bigr)^k}
 \end{equation}
 and $\zeta\equiv {\rm e}^{2\pi\sqrt{-1}/n}$.
\end{Theorem}

We need the assumptions on the eigenvalues and zeros of
the polynomial~\eqref{pol-} to display the expression in
the concise form. For Hamming graphs $H(d,2)$, $d\ge 2$
(or the hypercube of dimension $d$), we can obtain more
explicit results without such assumptions. This is
because the algebraic forms of the zeros of the quadratic
polynomial~\eqref{pol-} are available. In this sense,
we cannot expect to have general and explicit expressions
if $n\ge 7$.
This is because we need zeros of the polynomial of degree
$(n-1)$ (the unity is always a root of~\eqref{pol-}), and
if $n\ge 7$, we need explicit expressions of the roots of
the polynomial of degree larger than~5.

\begin{Corollary}[spectral representation of wave vector, $n=2$]
 \label{coro:n=2}
 Consider a random walk on the Hamming graph $H(d,2)$, $d\ge 2$
 satisfying assumption~\eqref{assump:1}. The wave vector of
 the quantum walks with coin operator~\eqref{coin} and
 the initial state~\eqref{init} is represented as
 \begin{align}
 \psi_{y,x}(t)=
 \frac{1}{2^d}\sqrt{\frac{w_{|y|}}{\kappa_{|y|}}}
 \Biggl\{&1+
 \frac{1}{2}\sum_{j\colon |\rho_j|<1}
 \Biggl[\frac{\bigl(\mu_j^+\bigr)^t}{1-\rho_j\mu_j^+}
 +\frac{\bigl(\mu_j^-\bigr)^t}{1-\rho_j\mu_j^-}\Biggr]K_j(|x|)
 \nonumber\\
 &{-}\,\frac{1}{2}\sum_{j\colon |\rho_j|<1}
 \Biggl[\frac{\bigl(\mu_j^+\bigr)^{t+1}}{1-\rho_j\mu_j^+}
 +\frac{\bigl(\mu_j^-\bigr)^{t+1}}{1-\rho_j\mu_j^-}\Biggr]
 K_j(|x\oplus y|)\nonumber\\
 &
 {+}\,\sum_{j>0\colon \rho_j=1}[(1-t)K_j(|x|)+tK_j(|x\oplus y|)]
 \nonumber\\
 &
 {+}\,\sum_{j>0\colon \rho_j=-1}(-1)^t[(1-t)K_j(|x|)-tK_j(|x\oplus y|)]
 \Biggr\} \label{spec_q2}
 \end{align}
 where
 \[
 \mu_j^{\pm}=\rho_j\pm \sqrt{-1}\sqrt{1-\rho_j^2},
 \qquad j\in\{0,\dots,d\},
 \]
 and $(\rho_j)$ are the eigenvalues of the random walk~\eqref{eigen}.
\end{Corollary}

We prepare a proposition to prove Theorem~\ref{theo:n>2}.

\begin{Proposition}\label{prop:ftr}
 Fix $\xi\in X\setminus\{0\}$ and assume
 $-1/(n-1)<\rho_{|\xi|}<1$ for prime $n$. Let
 \smash{$\mu^{(1)}_{|\xi|},\dots,\mu^{(n)}_{|\xi|}$} be the zeros
 of the polynomial~\eqref{pol-}. If they are distinct,
 the solution of system~\eqref{trans_qd} is represented as
 \begin{equation}\label{ftr}
 \tilde{\psi}_{y,\xi}(t)=
 \Biggl[
 \bigl(-\zeta^{-y\cdot\xi}\bigr)^t
 \Biggl(1-\sum_{i=1}^n
 \frac{2c_{|\xi|}^{(i)}}{1+\zeta^{y\cdot\xi}\mu_{|\xi|}^{(i)}}
 \Biggr)
 +\sum_{i=1}^n
 \frac{2c_{|\xi|}^{(i)}\bigl(\mu_{|\xi|}^{(i)}\bigr)^t}
 {1+\zeta^{y\cdot\xi}\mu^{(i)}_{|\xi|}}
 \Biggr]
 \tilde{\psi}_{y,\xi}(0), \qquad t\ge 0.
 \end{equation}
 In addition, \smash{{$\tilde{\psi}_{y,0}(t)=\tilde{\psi}_{y,0}(0)$}},
 $t\ge 0$.
\end{Proposition}

\begin{proof}
 Let
 \[
 \phi_{y,\xi}(t)=\sqrt{\frac{w_{|y|}}{\kappa_{|y|}}}
 \tilde{\psi}_{y,\xi}(t).
 \]
 Then, we recast~\eqref{trans_qd} into
 \begin{equation}\label{prop:ftr:1}
 \zeta^{-\xi\cdot y}\phi_{y,\xi}(t+1)=-\phi_{y,\xi}(t)
 +2\frac{w_{|y|}}{\kappa_{|y|}}\sum_{y'\in X}\phi_{y',\xi}(t),
 \qquad t\ge 0
 \end{equation}
 with the initial condition
 \[
 \phi_{y,\xi}(0)=\frac{w_{|y|}}{\kappa_{|y|}}\frac{1}{\sqrt{n^d}}.
 \]
 Summing up both sides of~\eqref{prop:ftr:1} in $y\in X$ gives
 \begin{gather}\label{prop:ftr:2}
 \sum_{y\in X}\zeta^{-y\cdot\xi}\phi_{y,\xi}(t+1)=
 -\sum_{y\in X}\!\phi_{y,\xi}(t)+2\sum_{y\in X}\!\frac{w_{|y|}}{\kappa_{|y|}}
 \sum_{y'\in X}\!\phi_{y',\xi}(t)
 =\sum_{y\in X}\!\phi_{y,\xi}(t),
 \qquad t\ge 0.\!\!\!
 \end{gather}
 On the other hand, multiplying $\zeta^{ky\cdot\xi}$,
 $k\in\{1,\dots,n-1\}$ and summing up both sides of~\eqref{prop:ftr:1} gives
 \begin{equation}\label{prop:ftr:3}
 \sum_{y\in X}\zeta^{(k-1)y\cdot\xi}\phi_{y,\xi}(t+1)=
 -\sum_{y\in X}\zeta^{ky\cdot\xi}\phi_{y,\xi}(t)
 +2\rho_{|\xi|}\sum_{y\in X}\phi_{y,\xi}(t),
 \qquad t\ge 0,
 \end{equation}
 where we used~\eqref{eigen_ftr}. Since $\zeta^{ny\cdot\xi}=1$,
 recursive use of~\eqref{prop:ftr:3} gives
 \begin{align*}
 \sum_{y\in X}\phi_{y,\xi}(t)&=
 -\sum_{y\in X}\zeta^{y\cdot\xi}\phi_{y,\xi}(t-1)
 +2\rho_{|\xi|}\sum_{y\in X}\phi_{y,\xi}(t-1)\\
 &=\sum_{y\in X}\zeta^{2y\cdot\xi}\phi_{y,\xi}(t-2)
 -2\rho_{|\xi|}\sum_{y\in X}\phi_{y,\xi}(t-2)
 +2\rho_{|\xi|}\sum_{y\in X}\phi_{y,\xi}(t-1)\\
 &=(-1)^{n-1}\sum_{y\in X}\zeta^{(n-1)y\cdot\xi}\phi_{y,\xi}(t-n+1)
 +2\rho_{|\xi|}\sum_{j=1}^{n-1}(-1)^{j-1}\sum_{y\in X}\phi_{y,\xi}(t-j)\\
 &=(-1)^{n-1}\sum_{y\in X}\phi_{y,\xi}(t-n)
 +2\rho_{|\xi|}\sum_{j=1}^{n-1}(-1)^{j-1}\sum_{y\in X}\phi_{y,\xi}(t-j),
 \qquad t\ge n,
 \end{align*}
 where~\eqref{prop:ftr:2} is used in the last equality.
 The recurrence relation for
 \[
 a_{|\xi|}(t):=\sum_{y\in X}\phi_{y,\xi}(t)
 \]
 is then
 \begin{equation}\label{prop:ftr:4}
 a_{|\xi|}(t)=2\rho_{|\xi|}
 \sum_{j=1}^{n-1}(-1)^{j-1}a_{|\xi|}(t-j)+(-1)^{n-1}a_{|\xi|}(t-n),
 \qquad t\ge n
 \end{equation}
 with the initial condition
 \begin{equation}\label{init_pol}
 a_{|\xi|}(t)=\frac{1}{\sqrt{n^d}}\bigl(2\rho_{|\xi|}-1\bigr)^{t-1}\rho_{|\xi|},
 \qquad n>t\ge 1, \qquad a_{|\xi|}(0)=\frac{1}{\sqrt{n^d}}.
 \end{equation}
 The characteristic polynomial of the recurrence relation~\eqref{prop:ftr:4} is~\eqref{pol-}.
 The zeros of the characteristic polynomials are denoted
 by \smash{$\mu_{|\xi|}^{(i)}$}, $i\in\{1,\dots,n\}$, where
 \smash{$\bigl|\mu_{|\xi|}^{(i)}\bigr|=1$}, $i\in\{1,\dots,n\}$ by~Lemma~\ref{lemm:pol}.
 Moreover, by Proposition~\ref{prop:rho}, they are not
 the negative of the roots of unity~$-\zeta^k$, ${k\in\mathbb{Z}}$.
 The solution of~\eqref{prop:ftr:4} is expressed as
 \begin{equation}\label{prop:ftr:5}
 a_{|\xi|}(t)=\frac{1}{\sqrt{n^d}}
 \sum_{i=1}^n c_{|\xi|}^{(i)}\bigl(\mu^{(i)}_{|\xi|}\bigr)^t,
 \qquad t\ge 0
 \end{equation}
 for some constants \smash{$c_{|\xi|}^{(1)},\dots,c_{|\xi|}^{(n)}$},
 \smash{$\sum_{i=1}^n c_{|\xi|}^{(i)}=1$}.
 Finding these constants is equivalent to finding the interpolating
 polynomial satisfying~\eqref{init_pol} at $t=0,1,\dots,n-1$.
 Namely, we have the matrix equation for \smash{$c^{(1)},\dots,c^{(n)}$}:
 \[
 \begin{pmatrix}
 1 & 1 & \cdots & 1\\
 \mu^{(1)} & \mu^{(2)} & \cdots & \mu^{(n)}\\
 \bigl(\mu^{(1)}\bigr)^2 & \bigl(\mu^{(2)}\bigr)^2 & \cdots
 & \bigl(\mu^{(n)}\bigr)^2\\
 \vdots & \vdots & \vdots & \vdots\\
 \bigl(\mu^{(1)}\bigr)^{n-1} & \bigl(\mu^{(2)}\bigr)^{n-1} & \cdots
 & \bigl(\mu^{(n)}\bigr)^{n-1}\\
 \end{pmatrix}
 \begin{pmatrix} c^{(1)}\\c^{(2)}\\c^{(3)}\\\vdots\\c^{(n)}
 \end{pmatrix}=
 \begin{pmatrix}1\\\rho\\(2\rho-1)\rho\\\vdots\\(2\rho-1)^{n-2}\rho
 \end{pmatrix},
 \]
 where the subfix $|\xi|$ is omitted for simplicity.
 The inverse of the Vandermonde matrix in the left-hand side has
 components
 \[
 \bigl[\mu^{j-1}\bigr] \frac{p(\mu)}{\bigl(\mu-\mu^{(i)}\bigr)p'\bigl(\mu^{(i)}\bigr)},
 \qquad i,j\in\{1,2,\dots,n\},
 \]
 where $\bigl[\mu^{j-1}\bigr]f(\mu)$ represents the coefficient of
 $\mu^{j-1}$ of the polynomial $f(\mu)$, $p(\mu)$ is the polynomial~\eqref{pol-},
 and $p'(\mu)$ is its derivative. The solution is~\eqref{coeff}.
 Substituting~\eqref{prop:ftr:5} into~\eqref{prop:ftr:1}~yields%
 \begin{align*}
 \phi_{y,\xi}(t)&=
 \frac{1}{\sqrt{n^d}}\frac{w_{|y|}}{\kappa_{|y|}}
 \Biggl(
 \bigl(-\zeta^{-y\cdot\xi}\bigr)^t
 -2\sum_{i=1}^nc_{|\xi|}^{(i)}\bigl(\mu^{(i)}_{|\xi|}\bigr)^t
 \sum_{j=1}^t\bigl(-\zeta^{y\cdot\xi}\mu^{(i)}_{|\xi|}\bigr)^{-j}
 \Biggr)\\
 &=\frac{1}{\sqrt{n^d}}\frac{w_{|y|}}{\kappa_{|y|}}
 \Biggl(
 \bigl(-\zeta^{-y\cdot\xi}\bigr)^t
 +2\sum_{i=1}^nc_{|\xi|}^{(i)}
 \frac{\bigl(\mu_{|\xi|}^{(i)}\bigr)^t-\bigl(-\zeta^{-y\cdot\xi}\bigr)^t}
 {1+\zeta^{y\cdot\xi}\mu^{(i)}_{|\xi|}}
 \Biggr)\\
 &=\frac{1}{\sqrt{n^d}}\frac{w_{|y|}}{\kappa_{|y|}}
 \Biggl[
 \bigl(-\zeta^{-y\cdot\xi}\bigr)^t
 \Biggl(1-\sum_{i=1}^n\frac{2c_{|\xi|}^{(i)}}{1+\zeta^{y\cdot\xi}\mu_{|\xi|}^{(i)}}\Biggr)
 +\sum_{i=1}^n
 \frac{2c_{|\xi|}^{(i)}\bigl(\mu_{|\xi|}^{(i)}\bigr)^t}
 {1+\zeta^{y\cdot\xi}\mu^{(i)}_{|\xi|}}
 \Biggr].
 \end{align*}
 In the second equality, we used the fact that
 \smash{$\mu_{|\xi|}^{(i)}\neq -\zeta^k$}, $k\in\mathbb{Z}$.
 The assertion \smash{$\tilde{\psi}_{y,0}(t)=\tilde{\psi}_{y,0}(0)$}
 immediately follows by~\eqref{prop:ftr:1} and~\eqref{prop:ftr:2}.
\end{proof}

\begin{proof}[Proof of Theorem~\ref{theo:n>2}]
 Since
 \[
 \sum_{l=0}^{n-1}\zeta^{l(a+b)}=n\delta_{a\oplus b,0},
 \qquad a,b\in\mathbb{Z},
 \]
 we recast~\eqref{ftr} into
 \begin{align*}
 \tilde{\psi}_{y,\xi}(t)&=
 \sqrt{\frac{w_{|y|}}{n^d\kappa_{|y|}}}
 \sum_{k=0}^{n-1}\sum_{l=0}^{n-1}\frac{\zeta^{l(k-y\cdot\xi)}}{n}
 \Biggl[\bigl(-\zeta^{-k}\bigr)^t
 \Biggl(1-\sum_{i=1}^n\frac{2c_{|\xi|}^{(i)}}{1+\zeta^{k}
 \mu_{|\xi|}^{(i)}}\Biggr)
 +\sum_{i=1}^n
 \frac{2c_{|\xi|}^{(i)}\bigl(\mu_{|\xi|}^{(i)}\bigr)^t}
 {1+\zeta^{k}\mu^{(i)}_{|\xi|}}
 \Biggr].
 \end{align*}
 The inside of the square brackets depends on $\xi$ through
 $|\xi|=\partial(0,\xi)$, \eqref{sum2} yields
 \begin{align*}
 {\psi}_{y,x}(t)={}&
 \frac{1}{n^d}\sqrt{\frac{w_{|y|}}{\kappa_{|y|}}}
 \sum_{k=0}^{n-1}\sum_{l=0}^{n-1}
 \frac{\zeta^{lk}}{n}
 \Biggl\{1+\sum_{\xi\in X\setminus\{0\}}\zeta^{-(x\oplus ly)\cdot\xi}
 \\
 &\hphantom{\frac{1}{n^d}\sqrt{\frac{w_{|y|}}{\kappa_{|y|}}}\sum_{k=0}^{n-1}\sum_{l=0}^{n-1}\frac{\zeta^{lk}}{n}\Biggl\{}{}
 {\times}
 \Biggl[\bigl(-\zeta^{-k}\bigr)^t
 \Biggl(1-\sum_{i=1}^n\frac{2c_{|\xi|}^{(i)}}{1+\zeta^{k}
 \mu_{|\xi|}^{(i)}}\Biggr)
 +\sum_{i=1}^n
 \frac{2c_{|\xi|}^{(i)}\bigl(\mu_{|\xi|}^{(i)}\bigr)^t}
 {1+\zeta^{k}\mu^{(i)}_{|\xi|}}
 \Biggr]\Biggr\}\\
 ={}&
 \frac{1}{n^d}\sqrt{\frac{w_{|y|}}{\kappa_{|y|}}}
 \sum_{l=0}^{n-1}\sum_{k=0}^{n-1}
 \frac{\zeta^{lk}}{n}
 \Biggl\{1+\sum_{j=1}^d K_j(|x\oplus ly|)\\
 &\hphantom{\frac{1}{n^d}\sqrt{\frac{w_{|y|}}{\kappa_{|y|}}}\sum_{l=0}^{n-1}\sum_{k=0}^{n-1}\frac{\zeta^{lk}}{n}\Biggl\{}{}
 {\times}
 \Biggl[\bigl(-\zeta^{-k}\bigr)^t
 \Biggl(1-\sum_{i=1}^n\frac{2c_{j}^{(i)}}{1+\zeta^{k}
 \mu_{j}^{(i)}}\Biggr)
 +\sum_{i=1}^n
 \frac{2c_{j}^{(i)}\bigl(\mu_{j}^{(i)}\bigr)^t}
 {1+\zeta^{k}\mu^{(i)}_{j}}
 \Biggr]\Biggr\}.
\!\!\!\!\!\!\tag*{\qed}
\end{align*}
\renewcommand{\qed}{}
\end{proof}

\begin{proof}[Proof of Corollary~\ref{coro:n=2}]
 The contribution from $\rho_0=1$ gives unity
 in the curly brackets, as the last assertion of
 Proposition~\ref{prop:ftr}.
 We begin with the cases with $|\rho_{|\xi|}|<1$.
 The recurrence relation~\eqref{prop:ftr:4} is
 \begin{align*}
 &a_{|\xi|}(t)-2\rho_{|\xi|}a_{|\xi|}(t-1)+a_{|\xi|}(t-2)=0, \qquad
 t\ge 2\\
 &\text{with} \ a_{|\xi|}(1)=\frac{\rho_{|\xi|}}{\sqrt{2^d}}, \
 a_{|\xi|}(0)=\frac{1}{\sqrt{2^d}}.
 \end{align*}
 The two zeros of the characteristic polynomial
 $x^2-2\rho_{|\xi|}x+1$, denoted by \smash{$\mu^+_{|\xi|}$} and
 \smash{$\mu^-_{|\xi|}$}, are distinct. We have
 \begin{equation}\label{coro:n=2:1}
 a_{|\xi|}(t)=\frac{\bigl(\mu_{|\xi|}^+\bigr)^t+\bigl(\mu_{|\xi|}^-\bigr)^t}{2\sqrt{2^d}},
 \qquad t\ge 0.
 \end{equation}
 Substituting~\eqref{coro:n=2:1} into~\eqref{ftr} gives
 \[
 \tilde{\psi}_{y,\xi}(t)=
 \Biggl\{\frac{\bigl(\mu^+_{|\xi|}\bigr)^t}
 {1+(-1)^{y\cdot\xi}\mu^+_{|\xi|}}
 +\frac{\bigl(\mu^-_{|\xi|}\bigr)^t}
 {1+(-1)^{y\cdot\xi}\mu^-_{|\xi|}}\Biggr\}
 \tilde{\psi}_{y,\xi}(0).
 \]
 It is convenient to work out for each cases of
 $z\cdot\xi$ is even or odd:
 \begin{align*}
 \tilde{\psi}_{y,\xi}(t)=
 &\Biggl\{
 \frac{1+(-1)^{y\cdot\xi}}{2}
 \Biggl[\frac{\bigl(\mu_{|\xi|}^+\bigr)^t}{1+\mu_{|\xi|}^+}
 +\frac{\bigl(\mu_{|\xi|}^-\bigr)^t}{1+\mu_{|\xi|}^-}\Biggr]
 +\frac{1-(-1)^{z\cdot\xi}}{2}
 \Biggl[\frac{\bigl(\mu_{|\xi|}^+\bigr)^t}{1-\mu_{|\xi|}^+}
 +\frac{\bigl(\mu_{|\xi|}^-\bigr)^t}{1-\mu_{|\xi|}^-}\Biggr]
 \Biggr\}\tilde{\psi}_{y,\xi}(0).
 \end{align*}
 This contributes to the inverse Fourier transform as
 \begin{align*}
 &\frac{1}{2^{d+1}}
 \sqrt{\frac{w_{|y|}}{\kappa_{|y|}}}
 \sum_{\xi\in\Xi}\Biggl\{
 \Biggl[
 \frac{\bigl(\mu_{|\xi|}^+\bigr)^t}{1-\rho_{|\xi|}\mu_{|\xi|}^+}
 +\frac{\bigl(\mu_{|\xi|}^-\bigr)^t}{1-\rho_{|\xi|}\mu_{|\xi|}^-}
 \Biggr](-1)^{-x\cdot\xi}
 \\
 &\hphantom{\frac{1}{2^{d+1}}\sqrt{\frac{w_{|y|}}{\kappa_{|y|}}}\sum_{\xi\in\Xi}\Biggl\{}{}
 -\Biggl[
 \frac{\bigl(\mu_{|\xi|}^+\bigr)^{t+1}}{1-\rho_{|\xi|}\mu_{|\xi|}^+}
 +\frac{\bigl(\mu_{|\xi|}^-\bigr)^{t+1}}{1-\rho_{|\xi|}\mu_{|\xi|}^-}
 \Biggr](-1)^{-(x\oplus y)\cdot\xi}
 \Biggr\}\\
 &\qquad{}=\frac{1}{2^{d+1}}
 \sqrt{\frac{w_{|y|}}{\kappa_{|y|}}}
 \sum_{j=0}^d
 \Biggl\{
 \Biggl[
 \frac{\bigl(\mu_{j}^+\bigr)^t}{1-\rho_{j}\mu_{j}^+}
 +\frac{\bigl(\mu_{j}^-\bigr)^t}{1-\rho_{j}\mu_{j}^-}
 \Biggr]K_{j}(|x|)
 \\
 &\hphantom{\qquad{}=\frac{1}{2^{d+1}} \sqrt{\frac{w_{|y|}}{\kappa_{|y|}}}\sum_{j=0}^d\Biggl\{}{}
 -\Biggl[
 \frac{\bigl(\mu_{j}^+\bigr)^{t+1}}{1-\rho_{j}\mu_{j}^+}
 +\frac{\bigl(\mu_{j}^-\bigr)^{t+1}}{1-\rho_{j}\mu_{j}^-}
 \Biggr]K_{j}(|x\oplus y|)
 \Biggr\},
 \end{align*}
 where the last equality follows by Lemma~\ref{lemm:sum}.
 For the cases with $\rho_{|\xi|}=\pm 1$, we have
 \[
 \tilde{\psi}_{y,\xi}(t)=
 \tilde{\psi}_{y,\xi}(0)
 \begin{cases}
 1,&\rho_{|\xi|}=1,\\ (1-2t)(-1)^t,&\rho_{|\xi|}=-1,
 \end{cases}
 \]
 for even $y\cdot\xi$ and
 \[
 \tilde{\psi}_{y,\xi}(t)=
 \tilde{\psi}_{y,\xi}(t)
 \begin{cases}
 1-2t,&\rho_{|\xi|}=1,\\(-1)^t,&\rho_{|\xi|}=-1.\end{cases}
 \]
 for odd $y\cdot\xi$. These expressions provide the two last
 lines of~\eqref{spec_q2}. Summing up all the contributions, we
 establish the assertion.
\end{proof}

\section{Limit distributions}
\label{sect:limit}

Since the probability function $P_t(x)$, $x\in X$ does not
converge as $t\to\infty$, Aharonov et al.~\cite{AAKV01} defined
the limit distribution of a quantum walk as the average over
infinitely long time
\[
 \bar{P}(x)=\lim_{T\to\infty}
 \frac{1}{T}\sum_{t=0}^{T-1} P_t(x)
 =\lim_{T\to\infty}
 \frac{1}{T}\sum_{t=0}^{T-1}\sum_{y\in X}|\psi_{y,x}(t)|^2.
\]
Intuitively, this quantity captures the proportion of time
which the quantum walker spends in state $x$.
In the following calculations, we use
\[
 \lim_{T\to\infty}\frac{1}{T}\sum_{t=0}^{T-1}
 {\rm e}^{\sqrt{-1} z t}=\delta_{z,0}, \qquad z\in\mathbb{C}.
\]

\subsection[Hamming graphs H(d,2) (hypercube)]{Hamming graphs $\boldsymbol{H(d,2)}$ (hypercube)}

For Hamming graphs $H(d,2)$, we have seen that~\eqref{spec_q2}
gives an explicit expression of a spectral representation of
the wave vector of the quantum walks. Suppose the eigenvalues of
the random walk satisfy $|\rho_j|<1$, $j\in\{1,\dots,d\}$
are distinct, and the eigenvalues of the evolution operator~$\mu^+_{j}$
and~$\mu^-_{j}$, $j\in\{1,\dots,d\}$ are distinct.
The mixture of i.i.d.\ updates for each coordinate
(Example~\ref{exam:miid}) for generic distribution of $\alpha$ is
an example satisfying this assumption. Then, the limit distribution
is
\begin{align}
\bar{P}(x)&=
\frac{1}{4^d}+\frac{1}{2\cdot 4^d}
\sum_{y\in X}\frac{w_{|y|}}{\kappa_{|y|}}
 \Biggl\{
 \sum_{j=1}^d
 \Biggl[
 \frac{\{K_j(|x|)\}^2}{1-\rho_j^2}+
 \frac{\{K_j(|x\oplus y|)\}^2}{1-\rho_j^2}\nonumber\\
 &\hphantom{=\frac{1}{4^d}+\frac{1}{2\cdot 4^d}\sum_{y\in X}\frac{w_{|y|}}{\kappa_{|y|}}\Biggl\{}{}
 -\frac{2\rho_j}{1-\rho_j^2}K_j(|x|)K_j(|x\oplus y|)
 \Biggr]\Biggr\}\nonumber\\
 &=\frac{1}{4^d}+\frac{1}{2\cdot 4^d}\sum_{j=1}^d
 \{K_j(|x|)\}^2
 +\frac{1}{2\cdot 4^d}\sum_{y\in X}\frac{w_{|y|}}{\kappa_{|y|}}
 \sum_{j=1}^d
 \frac{\{\rho_jK_j(|x|)-K_j(|x\oplus y|)\}^2}{1-\rho_j^2}
 \nonumber\\
 &=\frac{\begin{pmatrix}2(d-|x|)\\d-|x|\end{pmatrix}
 \begin{pmatrix}2|x|\\|x|\end{pmatrix}}
 {2\cdot 4^d\begin{pmatrix}d\\|x|\end{pmatrix}}
 +\frac{1}{2\cdot 4^d}
 \Biggl\{1+\sum_{y\in X}\frac{w_{|y|}}{\kappa_{|y|}}
 \sum_{j=1}^d
 \frac{\{\rho_jK_j(|x|)-K_j(|x\oplus y|)\}^2}{1-\rho_j^2}\Biggr\},
 \label{limit_q2}
\end{align}
where we used \cite[Theorem~3.1.3]{FF96}
\begin{equation}\label{identity_K}
 \sum_{j=0}^d \{K_j(|x|)\}^2=
 \frac{\begin{pmatrix}2(d-|x|)\\d-|x|\end{pmatrix}
 \begin{pmatrix}2|x|\\|x|\end{pmatrix}}
 {\begin{pmatrix}d\\|x|\end{pmatrix}}.
\end{equation}
The expression~\eqref{limit_q2} has an interpretation: the limit
distribution is the half-and-half mixture of the discrete arcsine law
and another probability distribution, because
\[
 \frac{1}{4^d}\begin{pmatrix}2(d-|x|)\\d-|x|\end{pmatrix}
 \begin{pmatrix}2|x|\\|x|\end{pmatrix}, \qquad |x|\in\{0,1,\dots,d\}
\]
is the probability mass function of the discrete arcsine law.
The limit distribution~\eqref{limit_q2} is has a~symmetry,
\begin{equation}\label{sym}
 \bar{P}(x)=\bar{P}(1-x),
\end{equation}
because if $n=2$ the identity $K_j(|1-x|)=K_j(d-|x|)=(-1)^j(|x|)$
follows by~\eqref{Kpol}.

\begin{Example}[the simple quantum walk]\label{exam:simple_q}
 The simple random walk was introduced in
 Example~\ref{exam:simple}. All the eigenvalues of
 the random walk are distinct and $\rho_d=-1$. For the wave
 vector of the quantum walk, the last line of~\eqref{spec_q2} yields
 \begin{align*}
 &\frac{1}{2^d}\sqrt{\frac{1}{d}}
 (-1)^t[(1-t)K_d(|x|)-tK_d(|x\oplus y|)]\\
 &\qquad{}=
 \frac{1}{2^d}\sqrt{\frac{1}{d}}
 (-1)^t\bigl[(1-t)(-1)^{|x|}-t(-1)^{|x\oplus y|}\bigr]=\frac{1}{2^d}(-1)^{t+|x|},
 \end{align*}
 since $|y|=1$. The limit distribution is
\begin{align*}
\bar{P}(x)
 &=\frac{\begin{pmatrix}2(d-|x|)\\d-|x|\end{pmatrix}
 \begin{pmatrix}2|x|\\|x|\end{pmatrix}}
 {2\cdot 4^d\begin{pmatrix}d\\|x|\end{pmatrix}}\nonumber
 +\frac{1}{4^d}
 +\frac{1}{2d\cdot 4^d}
 \sum_{j=1}^{d-1}
 \frac{\sum_{y\colon |y|=1}\{\rho_jK_j(|x|)-K_j(|x\oplus y|)\}^2}{1-\rho_j^2}.
\end{align*}
This coincides with equation~13 in Ho et al.~\cite{HIKST18}
by the following identity.
This limit distribution has the symmetry~\eqref{sym}.
\end{Example}

\begin{Proposition}
 For $j\in\{1,\dots,d-1\}$ and $\rho_j=1-2j/d$,
 \begin{equation}\label{hikst}
 \sum_{z\colon |z|=1}
 \{\rho_jK_j(|x|)-K_j(|x\oplus z|)\}^2=
 \left(1-\frac{|x|}{d}\right)|x|
 \{K_j(|x|-1)-K_j(|x|+1)\}^2,
 \end{equation}
 where $x\in X$.
\end{Proposition}

\begin{proof}
 If $|x|=0$ or $d$, the equality holds because $\rho_jK_j(0)=K_j(1)$
 and $\rho_jK_j(d)=K_j(d-1)$, respectively.
 Otherwise, expanding the left-hand side of~\eqref {hikst} yields
 \begin{align*}
 &d\{\rho_jK_j(|x|)\}^2-2\rho_jK_j(|x|)\{(d-|x|)K_j(|x|+1)
 +|x|K_j(|x|-1)\}\\
 &\qquad{}+(d-|x|)\{K_j(|x|+1)\}^2+|x|\{K_j(|x|-1)\}^2.
 \end{align*}
 We recast this into the right-hand side of~\eqref{hikst} by
 expressing $K_j(|x|)$ with $K_j(|x|-1)$ and ${K_j(|x|+1)}$
 by using the three-term recurrence relation of the
 Krawtchouk polynomials \cite[equation~(1.10.3)]{KS98}:
 \[
 \frac{i}{d}K_j(i-1)+\left(1-\frac{i}{d}\right)K_j(i+1)
 =\left(1-\frac{2j}{d}\right)K_j(i), \qquad i\in\{1,\dots,d-1\}.
\tag*{\qed}
\]
\renewcommand{\qed}{}
\end{proof}

\begin{Example}[the independent quantum walk]\label{exam:indep_q}
 The independent random walk was introduced in
 Example~\ref{exam:indep}.
 The eigenvalues of this random walk degenerate.
 For the quantum walk, the wave
 vector is
 \begin{align*}
 \psi_{y,x}(t)=\frac{1}{2^{3d/2}}
 \biggl\{&1+\frac{1}{2}\bigl[\bigl(\sqrt{-1}\bigr)^t+\bigl(-\sqrt{-1}\bigr)^t\bigr]\bigl(2^d\delta_{|x|,0}-1\bigr) \\
 & -\frac{1}{2}\bigl[\bigl(\sqrt{-1}\bigr)^{t+1}+\bigl(-\sqrt{-1}\bigr)^{t+1}\bigr]
 \bigl(2^d\delta_{|x\oplus y|,0}-1\bigr)\biggr\}.
 \end{align*}
 The limit distribution is
 \[
 \bar{P}(x)
 =\biggl(\frac{1}{2}-\frac{1}{2^d}\biggr)\delta_{x,0}
 +\frac{1}{2\cdot 2^{d}}+\frac{1}{4^d}.
 \]
 As $d\to\infty$, this is the half-and-half mixture of
 the atom at the origin and the uniform distribution.
 The limit distribution does not have the symmetry~\eqref{sym}.
\end{Example}

\begin{Example}[the non-local quantum walk]
 The non-local random walk was introduced in
 Example~\ref{exam:nlrw}. To make the expressions
 explicit, let the cardinality $m=2$ and assume $d\ge 3$
 is odd. Then,
 \[
 \rho_i=\frac{K_2(i)}{\kappa_2}=1-\frac{4i}{d-1}+\frac{4i^2}{d(d-1)}
 >-\frac{1}{d-1}, \qquad i\in\{0,\dots,d\},
 \]
 where $\rho_i=\rho_{d-i}$, $i\in\{0,\dots,d\}$.
 For the wave vector of the quantum walk, the second last line
 of~\eqref{limit_q2} yields
 \begin{align*}
 &\frac{1}{2^d}\sqrt{\frac{2}{d(d-1)}}
 [(1-t)K_d(|x|)-tK_d(|x\oplus y|)]\\
 &\qquad{}=
 \frac{1}{2^d}\sqrt{\frac{2}{d(d-1)}}
 \bigl[(1-t)(-1)^{|x|}-t(-1)^{|x\oplus y|}\bigr]
 =\frac{1}{2^d}\sqrt{\frac{2}{d(d-1)}}(-1)^{|x|},
 \end{align*}
 since $|y|=2$. The limit distribution is
\begin{align*}
\bar{P}(x)
 ={}&\frac{\begin{pmatrix}2(d-|x|)\\d-|x|\end{pmatrix}
 \begin{pmatrix}2|x|\\|x|\end{pmatrix}}
 {2\cdot 4^d\begin{pmatrix}d\\|x|\end{pmatrix}}\nonumber
 +\frac{1}{4^d}\\
 &{+}\,\frac{1}{4^d}
 \frac{2}{d(d-1)}\sum_{j=1}^{(d-1)/2}
 \frac{\sum_{y\colon |y|=2}\{\rho_jK_j(|x|)-K_j(|x\oplus y|)\}^2}{1-\rho_j^2},
\end{align*}
which has the symmetry~\eqref{sym}.
\end{Example}

\begin{Example}[the mixture of i.i.d.\ updates for each coordinate]
 The limit distribution is the mixture of~\eqref{limit_q2}
 with the distribution of parameter $\alpha\in(0,1)$.
 If it has the single atom at $r\in(0,1)\setminus\{1/2\}$, we have
\begin{align}
\bar{P}(x)={}&\frac{\begin{pmatrix}2(d-|x|)\\d-|x|\end{pmatrix}
 \begin{pmatrix}2|x|\\|x|\end{pmatrix}}
 {2\cdot 4^d\begin{pmatrix}d\\|x|\end{pmatrix}}
 +\frac{1}{2\cdot 4^d}\nonumber\\
 &{+}\,\frac{1}{2\cdot 4^d}
 \sum_{y\in X}r^{|y|}(1-r)^{d-|y|}
 \sum_{j=1}^d
 \frac{\{(1-2r)^jK_j(|x|)-K_j(|x\oplus y|)\}^2}{1-(1-2r)^j},
 \label{miid}
\end{align}
which has the symmetry~\eqref{sym}.
If $r=1/2$, this model reduces to Example~\ref{exam:indep_q},
where the eigenvalues of the random walk degenerate, while
\eqref{miid} reduces to the half-and-half mixture of the
discrete arcsine law and the uniform distribution, where
we used
\[
 \sum_{y\in X}\sum_{j=0}^d\{K_j(|y|)\}^2=
 \sum_{i=0}^d\begin{pmatrix}2(d-i)\\d-i\end{pmatrix}
 \begin{pmatrix}2i\\i\end{pmatrix}=4^d.
\]
Here, the first equality follows by~\eqref{identity_K},
and the second equality can be obtained by noting that%
\[
\sum_{i=0}^\infty\begin{pmatrix}2i\\i\end{pmatrix}s^i=
\frac{1}{\sqrt{1-4s}}.
\]
Figure~\ref{figure1} displays the limit distributions of quantum walks
for $d=10$. Independent quantum walk
(Example~\ref{exam:indep_q}) and i.i.d.\ updates with
a single atom at $r=0.1$, $r=0.49$, and $r=0.9$ are shown.
When $r$ is close to $0$ or $1$, large probability masses
are concentrated around $0$ and $(1,\dots,1)$.
These are similar to the simple quantum walks (Example~\ref{exam:simple_q})
in that they are concave in the center (see \cite[Figure~2]{MPAD08}).
If $r=0.49$, the limit distribution is close to the half-and-half
mixture of the discrete arcsine law and the uniform
distributions, but if $r=1/2$ (the independent quantum walk),
the symmetry~\eqref{sym} breaks down and
the component other than the uniform distribution is
piled up at the origin. This drastic change of the behaviour
at~${r=1/2}$ is reminiscent of the fact that the random walk
mixes in exactly one step at $r=1/2$.
\begin{figure}[!ht]
 \centering
 \includegraphics[width=.65\textwidth]{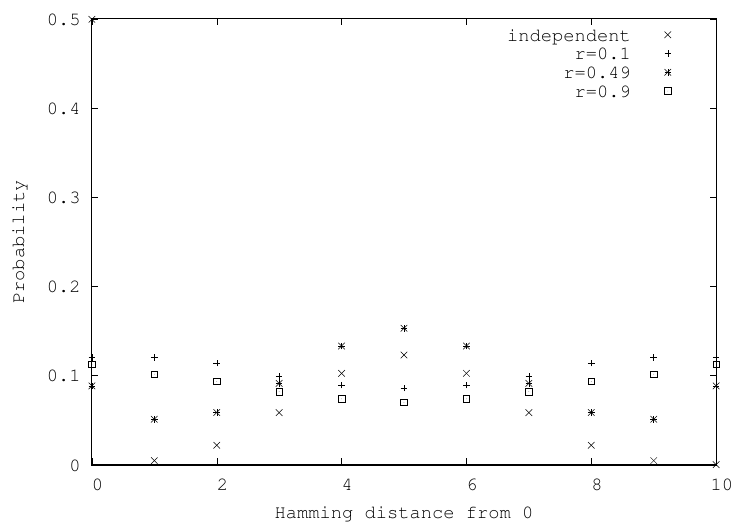}
 \caption{The limit distributions of some quantum walks.}\label{figure1}
 \end{figure}
\end{Example}

\subsection[Hamming graphs H(d,n) for prime n>=3]{Hamming graphs $\boldsymbol{H(d,n)}$ for prime $\boldsymbol{n\ge 3}$}

We prepare the following identities.

\begin{Proposition}
 For odd $n\ge 3$, we have
 \begin{align}
 &\sum_{k=0}^{n-1}\frac{1}{1+\zeta^k}=\frac{n}{2}, \label{prop:sum:1}\\
 &\sum_{k=0}^{n-1}\frac{\zeta^{lk}}{1+\zeta^k}=\frac{n}{2}(-1)^{l-1},
 \qquad l\in\{1,\dots,n-1\},\label{prop:sum:2}\\
 &\sum_{k=0}^{n-1}\frac{1}{\bigl(1+\zeta^k\bigr)\bigl(1+\bar{\zeta}^k\bigr)}=\frac{n^2}{4},
 \label{prop:sum:3}
 \end{align}
 where $\zeta={\rm e}^{2\pi\sqrt{-1}/n}$.
\end{Proposition}

\begin{proof}
 Note that
 \[
 \frac{2}{1+\zeta^k}=\frac{1-\bigl(-\zeta^k\bigr)^n}{1-\bigl(-\zeta^k\bigr)}
 =\sum_{i=0}^{n-1}\bigl(-\zeta^k\bigr)^i.
 \]
 For~\eqref{prop:sum:1}, we have
 \[
 \frac{1}{2}\sum_{k=0}^{n-1}\sum_{i=0}^{n-1}(-1)^i\zeta^{ki}
 =\frac{n}{2}+\frac{1}{2}\sum_{i=1}^{n-1}(-1)^i\sum_{k=0}^{n-1}\zeta^{ki}
 =\frac{n}{2}+\frac{1}{2}\sum_{i=1}^{n-1}(-1)^i
 \frac{1-\zeta^{in}}{1-\zeta^{i}}=\frac{n}{2}.
 \]
 For~\eqref{prop:sum:2}, we have
 \begin{align*}
\frac{1}{2}\sum_{k=0}^{n-1}\sum_{i=0}^{n-1}(-1)^i\zeta^{k(i+l)}
 &=\frac{n}{2}(-1)^{n-l}+\frac{1}{2}\sum_{i\in\{0,\dots,n-1\}\setminus\{n-l\}}
 (-1)^i\sum_{k=0}^{n-1}\zeta^{k(i+l)}\\
 &=\frac{n}{2}(-1)^{n-l}+\frac{1}{2}\sum_{i\in\{0,\dots,n-1\}\setminus\{n-l\}}
 (-1)^i\frac{1-\zeta^{(i+l)n}}{1-\zeta^{i+l}}=\frac{n}{2}(-1)^{l+1}.
 \end{align*}
 For~\eqref{prop:sum:3}, we have
 \begin{align*}
\sum_{k=0}^{n-1}\frac{1}{\bigl(1+\zeta^k\bigr)\bigl(1+\bar{\zeta}^k\bigr)}
 &{}=\frac{1}{4}\sum_{k=0}^{n-1}\sum_{i=0}^{n-1}\sum_{j=0}^{n-1}
 \bigl(-\zeta^k\bigr)^i\bigl(-\zeta^{-k}\bigr)^j=
 \frac{1}{4}\sum_{i=0}^{n-1}\sum_{j=0}^{n-1}
 \bigl(-\zeta^k\bigr)^{i-j}\\
 &{}=\frac{1}{4}\sum_{i=0}^{n-1}\sum_{j=0}^{n-1}(-1)^{i-j}
 \sum_{k=0}^{n-1}\bigl(\zeta^{i-j}\bigr)^k\\
 &{}=
 \frac{1}{4}\sum_{i=0}^{n-1}n+
 \frac{1}{4}\sum_{i=0}^{n-1}\sum_{j\in\{0,\dots,n-1\}\setminus\{i\}}
 (-1)^{i-j}\frac{1-\zeta^{(i-j)n}}{1-\zeta^{i-j}}=\frac{n^2}{4}.
\tag*{\qed}
\end{align*}
\renewcommand{\qed}{}
\end{proof}

With using these identities, we obtain the following example.

\begin{Example}[the independent quantum walk]
 The wave vector is
 \[
 \psi_{y,x}(t)=\frac{1}{n^{3d/2}}
 \sum_{l=0}^{n-1}\sum_{k=0}^{n-1}
 \frac{\zeta^{lk}}{n}
 \Biggl[1+\frac{2}{n}\bigl(n^d\delta_{|x\oplus ly|,0}-1\bigr)
 \sum_{i=0}^{n-1}
 \frac{\zeta^{it}}{1+\zeta^{ik}}\Biggr].
 \]
 The limit distribution is
 \[
 \bar{P}(x)=\biggl(1-\frac{1}{n}\biggr)\frac{1}{n^d}+\frac{2(n-1)}{n^{2d+1}}
 +\biggl(\frac{1}{n}-\frac{2(n-1)}{n^{d+1}}\biggr)\delta_{x,0}.
 \]
 As $d\to\infty$, this is the $(1-1/n)$-and-$1/n$ mixture of the
 uniform distribution and the atom at the origin.
\end{Example}

The spectral representations for $n\ge 3$ becomes much more
involved than those for $n=2$. For the simple random walk of
$n=3$, the eigenvalues are
\[
 \rho_0=1, \qquad \rho_d=-1/2, \qquad \text{and} \qquad
 \rho_j=1-\frac{3j}{2d}, \qquad j\in\{1,2,\dots,d-1\},
\]
where $\rho_j\in(-1/2,1)$, $j\in\{1,2,\dots,d-1\}$.
The corresponding eigenvalues of the evolution operator of
the quantum walk are
\begin{alignat*}{3}
 &1,\ -\zeta, \ -\zeta^2 \qquad && \text{for}\ \rho_0=1,&\\
 &1,\ -1 \qquad && \text{for}\ \rho_d=-1/2,&\\
 &1,\ {\rm e}^{\pm\sqrt{-1}\theta_j} \qquad && \text{for}\ \rho_j, ~ j\in\{1,2,\dots,d-1\},&
\end{alignat*}
where
\[
 \cos\theta_j=\rho_j-\frac{1}{2}=\frac{d-3j}{2d}.
\]
For $j\in\{1,2,\dots,d-1\}$,
\[
 c_j^{(1)}=\frac{d}{d+3j}, \qquad
 c_j^{(2)}=c_j^{(3)}=\frac{3j}{2(d+3j)}.
\]

Since $\rho_d=-1/2$ violates the assumption of Theorem~\ref{theo:n>2},
we consider the contribution separately. The same argument to
obtain Proposition~\ref{prop:ftr} yields
\[
 \tilde{\psi}_{y,\xi}(t)=\frac{1}{\sqrt{3^d\cdot 2d}}
 \sum_{k=1}^2\sum_{l=0}^2\frac{\zeta^{l(k-y\cdot\xi)}}{6}
 \biggl[
 \frac{1}{1+\zeta^k}-\frac{3(-1)^t}{1-\zeta^k}
 +\frac{2\bigl(2+2\zeta^k-\zeta^{2k}\bigr)}{1-\zeta^{2k}}\bigl(-\zeta^{3-k}\bigr)^t
 \biggr]
\]
for $|\xi|=d$, where we used the fact that $y\cdot\xi\neq 0$
if $|\xi|=d$ and $|y|=1$. Then, the wave vector is
\begin{align*}
&\psi_{y,x}(t)=
 \frac{1}{3^d\sqrt{2d}}
 \sum_{k=0}^2\sum_{l=0}^2\frac{\zeta^{lk}}{3}\\
 &\qquad{}{\times}\Biggl[1+\sum_{j=1}^{d-1}K_j(|x\oplus ly|)
 \Biggl(
 \frac{2c_j^{(1)}}{1+\zeta^k}+\frac{2c_j^{(2)}{\rm e}^{\sqrt{-1}\theta_j t}}
 {1+\zeta^k{\rm e}^{\sqrt{-1}\theta_j}}+
 \frac{2c_j^{(2)}{\rm e}^{-\sqrt{-1}\theta_j t}}{1+\zeta^k{\rm e}^{-\sqrt{-1}\theta_j}}
 \Biggr)\Biggr]\\
 &\qquad{}{+}\, \frac{1}{3^d\sqrt{2d}}
\sum_{k=1}^2\sum_{l=0}^2\frac{\zeta^{lk}}{6}K_d(|x\oplus ly|)
 \Biggl[
 \frac{1}{1+\zeta^k}
 -\frac{3(-1)^t}{1-\zeta^k}
 +\frac{2\bigl(2+2\zeta^k-\zeta^{2k}\bigr)}{1-\zeta^{2k}}\bigl(-\zeta^{3-k}\bigr)^t
 \Biggr].
\end{align*}
We have
\begin{align*}
\psi_{y,x}(t)={}&\frac{1}{3^d\sqrt{2d}}
 \Biggl\{1+\sum_{j=1}^{d-1}c_j^{(1)}
 [K_j(|x|)+K_j(|x\oplus y|)-K_j(|x\oplus 2y|)] \\
 &{+}\,\frac{1}{6}[K_d(|x|)+K_d(|x\oplus y|)-2K_d(|x\oplus 2y|)]
 -\frac{1}{2}[K_d(|x|)-K_d(|x\oplus y|](-1)^t\\
 &{+}\,\bigl[K_d(|x|)+\zeta K_d(|x\oplus y|)+\zeta^2 K_d(|x\oplus 2y|)\bigr]
 \frac{\bigl(-\zeta^2\bigr)^t}{1-\zeta}\\
 &{-}\,
 \bigl[\zeta K_d(|x|)+K_d(|x\oplus y|)+\zeta^2 K_d(|x\oplus 2y|)\bigr]
 \frac{(-\zeta)^t}{1-\zeta}\\
 &{+}\,\sum_{j=1}^{d-1}\bigl[K_j(|x|)+K_j(|x\oplus y|){\rm e}^{2\sqrt{-1}\theta_j}+
 K_j(|x\oplus 2y|){\rm e}^{\sqrt{-1}\theta_j}\bigr]
 \frac{2c_j^{(2)}{\rm e}^{\sqrt{-1}\theta_jt}}{1+{\rm e}^{3\sqrt{-1}\theta_j}}\\
 & {+}\,\sum_{j=1}^{d-1}\bigl[K_j(|x|)+K_j(|x\oplus y|){\rm e}^{-2\sqrt{-1}\theta_j}+
 K_j(|x\oplus 2y|){\rm e}^{-\sqrt{-1}\theta_j}\bigr]
 \frac{2c_j^{(2)}{\rm e}^{-\sqrt{-1}\theta_jt}}{1+{\rm e}^{-3\sqrt{-1}\theta_j}}
 \Biggr\}.
\end{align*}
The limit distribution is
\begin{align*}
\bar{P}(x)={}&\frac{1}{9^d}
 +\frac{4^{d-|x|}}{9^d}\frac{17d+45|x|}{8d}
 -\biggl(\frac{2}{9}\biggr)^d\biggl(-\frac{1}{2}\biggr)^{|x|+1}
 +\frac{2}{9^d}\sum_{j=1}^{d-1}\frac{K_j(|x|)}{1+3j/d}\\
 &{+}\,\frac{1}{2d\cdot9^d}
 \sum_{y\colon |y|=1}\Biggl[\sum_{j=1}^{d-1}
 \frac{K_j(|x|)+K_j(|x\oplus y|)-K_j(|x\oplus 2y|)}{1+3j/d}\Biggr]^2\\
 &{+}\,\frac{2^d}{12d\cdot9^d}\sum_{j=1}^{d-1}\sum_{y:|y|=1}
 \frac{K_j(|x|)+K_j(|x\oplus y|)-K_j(|x\oplus 2y|)}{1+3j/d}\\
 &\hphantom{{+}\,\frac{2^d}{12d\cdot9^d}\sum_{j=1}^{d-1}\sum_{y:|y|=1}}{}
 \times\Bigl[
 \begin{pmatrix}-\frac{1}{2}\end{pmatrix}^{|x|}
 +\begin{pmatrix}-\frac{1}{2}\end{pmatrix}^{|x\oplus y|}
 -2\begin{pmatrix}-\frac{1}{2}\end{pmatrix}^{|x\oplus 2y|}\Bigr]\\
 &{+}\,\frac{1}{3\cdot 9^d}
 \sum_{j=1}^{d-1}\frac{1}{d-j}\sum_{y:|y|=1}
 \Biggl|\frac{K_j(|x|)+K_j(|x\oplus y|){\rm e}^{2\sqrt{-1}\theta_j}+
 K_j(|x\oplus 2y|){\rm e}^{\sqrt{-1}\theta_j}}{1+3j/d}\Biggr|^2,
\end{align*}
where we used $K_d(j)=2^d(-1/2)^j$, $j\in\{0,1,\dots,d\}$ and
Table~\ref{table1}.

\begin{table}[!ht]\centering
\caption{$|x\oplus y|$ and $|x\oplus 2y|$ for $|y|=1$, $y_i\neq 0$.}\label{table1}\vspace{1mm}

\begin{tabular}{cccc}
 & & $|x\oplus y|$ & $|x\oplus 2y|$\\
 \hline
 $x_i=0$ & & $|x|+1$ & $|x|+1$\\
 $x_i=1$ & $y_i=1$ & $|x|$ & $|x|-1$\\
 $x_i=1$ & $y_i=2$ & $|x|-1$ & $|x|$\\
 $x_i=2$ & $y_i=1$ & $|x|-1$ & $|x|$\\
 $x_i=2$ & $y_i=2$ & $|x|$ & $|x|-1$\\
 \hline
\end{tabular}
\end{table}

\subsection*{Acknowledgements}

Shuhei Mano acknowledges the hospitality of School of
Mathematics, Monash University, where this work was started.
He was supported in part by JSPS KAKENHI Grants 18H00835 and
24K06876. The authors would like to thank the referees
for their careful reading and for providing useful comments.


\pdfbookmark[1]{References}{ref}
 \LastPageEnding

\end{document}